\documentclass[usenatbib]{mn2e}

\usepackage{graphicx}
\usepackage{subfigure}

\begin{document}

\title[Time-dependent radiative transfer]{An algorithm for Monte-Carlo time-dependent radiation transfer}
\author[Tim J. Harries]{Tim J. Harries\\
School of Physics, University of Exeter, Stocker Road, Exeter EX4
4QL. Email: th@astro.ex.ac.uk
}

\maketitle

\begin{abstract}

A new Monte-Carlo algorithm for calculating time-dependent radiative-transfer under the assumption of LTE is presented.  Unlike flux-limited
diffusion the method is polychromatic, includes scattering, and is
able to treat the optically thick and free-streaming regimes
simultaneously. The algorithm is tested on a variety of 1-d and 2-d
problems, and good agreement with benchmark solutions is found. The
method is used to calculate the time-varying spectral energy
distribution from a circumstellar disc illuminated by a protostar whose
accretion luminosity is varying. It is shown that the time lag between
the optical variability and the infrared variability results from a
combination of the photon travel time and the thermal response in the
disc, and that the lag is an approximately linear function of
wavelength.

\end{abstract}


\section{Introduction}

Theoretical astrophysics often necessitates computing the energy
balance between gas and an ambient radiation field. Most frequently
this involves solving the equation of radiative equilibrium, and a
wide variety of theoretical tools have been developed to tackle this
problem efficiently, ranging from those which make simplifying
assumptions both to the geometry (spherical or plane-parallel) and to
the underlying microphysics (the gray approximation for example)
through to a full poly-chromatic treatment in multiple
dimensions.

Although most radiation transfer (RT) codes assume thermal balance
many interesting phenomena occur out of equilibrium, and here a
time-dependent approach is required. A spectacular example of this
occurs during supernova explosions, and progress is being made in
time-dependent modelling of such objects
e.g. \cite{jack_2009, kromer_2009}. 

Time-dependent methods are also required for radiation hydrodynamics (RHD). Broadly speaking the transport methods can be split into those concerned with ionizing radiation, which generally adopt ray-casting methods (e.g. \citealt{dale_2007, maclow_2007}), and those that deal with transport in dusty media which use either short-characteristic methods, Monte-Carlo techniques, or the diffusion approximation (e.g. \citealt{hofner_2003, woitke_2006, turner_2001}). RHD codes may be further subdivided into those  which are applicable to situations in which the radiation transport/thermal equilibrium timescale for the gas is shorter than characteristic hydrodynamical timescale, and those in which the hydrodynamical timescale is the shorter. In the former case the radiation transport may be conducted as a sequence of pseudo-equilibrium steps (e.g. \citealt{hofner_2003, woitke_2006, dale_2007, maclow_2007, freytag_2008, acreman_2010}). However  when the thermal timescale becomes comparable to, or indeed larger than, the hydrodynamical timescale the time-dependent form of the radiation-transfer equations must be solved  (e.g. \citealt{krumholz_2007a}).

In many situations RT is implemented assuming energy transport occurs via radiative diffusion of photons. This is a good assumption for optically thick regions, but breaks down in the
optically thin regime, since the mean-free-path of photons (and thus
the speed of the diffusing radiation field) can become arbitrarily
large; a flux-limiter is adopted to control
this problem. Flux-limited diffusion (FLD) is normally
implemented in the gray approximation
\citep{turner_2001, whitehouse_2004, whitehouse_2005, krumholz_2007},
although comparisons with a polychromatic treatment demonstrate that
this is often quite a poor approximation
\citep{preibisch_1995}. Opaque obstacles also cause problems for FLD,
since the radiation field can diffuse around the obstacle when in fact
a shadow should be cast.

A modified version of the Monte Carlo (MC) radiative equilibrium method
developed by \cite{lucy_1999} offers an attractive route to full time-dependent RT,
since it would allow a full polychromatic treatment within which
multiple scattering may be incorporated. The method is computationally
demanding, but it  has the advantage of being extremely efficient
to parallelize. Indeed the feasibility of using such a method has
already been demonstrated, at least in the case where the thermal
timescales are short \citep{acreman_2010}.

Here I present an algorithm for calculating the matter and radiation
field energy densities, as a function of time, for an arbitrary
distribution photon of sources embedded in an arbitrary distribution of
gas  under LTE conditions. I describe  one-dimensional benchmark tests of the method,
and detail its incorporation into the {\sc torus} radiation-transfer
code. I test the new version of {\sc torus} against a two-dimensional
disc benchmark \citep{pascucci_2004}. Finally I give a simple
application of the code demonstrating the effect of a varying
accretion luminosity from a protostar on its surrounding dusty
circumstellar disc.

\section{Method}

We describe the physical quantities (densities, temperatures etc) on
a cell-centred grid. For a gas in LTE at temperature $T$ the rate at which it
emits energy is given by
\begin{equation}
\dot{E} = 4 \pi \int_0^\infty k_\nu B_\nu \, d\nu,
\label{eq:emissivity}
\end{equation}
where $k_\nu$ is the absorption coefficient and $B_\nu$ is the Planck
function. The rate at which the same gas absorbs energy is given by
\begin{equation}
\dot{A} = 4 \pi \int_0^\infty k_\nu J_\nu \, d\nu,
\label{eq:absorption-rate}
\end{equation}
where $J_\nu$ is the mean intensity of the radiation field. Clearly if
the gas is in radiative equilibrium then $\dot{A}=\dot{E}$ and we find
\begin{equation}
T = \left( \frac{\dot{A}}{4 \sigma \kappa_P} \right)^{1/4}
\label{eq:rad-eq-time}
\end{equation}
where $\kappa_P$ is the Planck-mean absorption coefficient. However if
we consider gas that is not in radiative equilibrium then the nett
change in energy density of the gas
\begin{equation}
\dot{u}_g = \dot{A} - \dot{E}
\label{eq:a-minus-e}
\end{equation}
and, {\it mutatis mutandis}, the rate of change in energy density of
the radiation field is 
\begin{equation}
\dot{u}_r = \dot{E} - \dot{A}
\label{eq:e-minus-a}
\end{equation}
Now we consider a gas of volume $V$ at time $t$. Within this volume is
a star of luminosity $L_*$.  The luminosity of the gas is given by
\begin{equation}
L_g = \int_V \dot{E} \, dV.
\end{equation}
We assume that the temperature of the gas is constant over a single
timestep $\Delta t$. During this timestep we assume that the gas and
the star produce $N_g$ and $N_*$ new  photon packets
respectively. The individual photon packet energies are given by
\begin{eqnarray}
\epsilon_g = \frac{L_g \Delta t}{N_g} & \nonumber & \epsilon_* =
\frac{L_* \Delta t}{N_*} \\
\end{eqnarray}
The energy density is related to the temperature by
\begin{equation}
u_g =  \frac{ R T \rho}{(\gamma-1) \mu}
\label{eq:gas-energy-density}
\end{equation}
where $R$ is the gas constant, $\rho$ is the mass density, $\gamma$ is
the ratio of specific heats and $\mu$ is the mean molecular weight. We
follow \cite{lucy_1999} and use the result that the energy density of
the radiation field in the interval $(\nu, \nu+d\nu)$ is given by
\begin{equation}
u_{r,\nu} = 4 \pi J_\nu d\nu / c.
\label{eq:energy-density}
\end{equation}
A photon packet moving between events (scatterings, absorptions,
crossing grid-cell boundaries) contributes an energy $\epsilon_\nu$ for a
time $\ell / c$ (where $\ell$ is the distance between events) to the
local energy density. The photon energy density is therefore
\begin{equation}
u_r = \frac{1}{\Delta t} \frac{1}{V} \frac{1}{c} \sum \epsilon_{\nu} \ell 
\label{eq:energy-summation}
\end{equation}
where the summation is over all photon packets. Now combining equations \ref{eq:absorption-rate} and
\ref{eq:energy-density} with equation \ref{eq:energy-summation} we
obtain an expression for the energy absorption rate:
\begin{equation}
\dot{A} = \frac{1}{V} \frac{1}{\Delta t} \sum k_\nu  \epsilon_\nu \ell 
\label{eq:new_absorption-rate}
\end{equation}
The new energy density of the gas may then be calculated
\begin{equation}
u^{n+1}_g = u^{n}_g + (\dot{A} - \dot{E}) \Delta t.
\label{eq:ug-update}
\end{equation}
This explicit integration scheme will require a careful choice of the
timestep $\Delta t$, which must be short enough to ensure stability
while also being long enough that the computation remains tractable.
Timescale considerations are discussed in section~\ref{section:timescales}.

It is worth noting that equation~\ref{eq:ug-update} is not the only method for
updating the internal energy of the gas. Since we follow the radiation
field in detail we can, for each cell, calculate the energy in photon
packets entering and leaving the cell, and hence the change in
radiation energy density due to the RT ($\Delta u_{r,{\rm trans}}$).
  MC estimators exist for $u_r$ for both the start and the end of the
  timestep, so using
\begin{equation}
u^{n+1}_r = u^n_r +  (\dot{E} - \dot{A}) \Delta t + \Delta u_{r,{\rm
    trans}}
\end{equation}
we can find $(\dot{E} - \dot{A}) \Delta t$ and substitute it
into equation~\ref{eq:ug-update}, bypassing the need to find the an
estimator for $\dot{A}$. In practise we found that this method
required a larger number of photon packets (in order to improve the
estimate of $u_r$ and $\Delta u_{r,{\rm trans}}$) than that needed using
  an estimator for $\dot{A}$.

A single timestep encompasses of a loop over photon packets, each with
an individual energy $\epsilon_\nu$ and frequency $\nu$. Information on
photon packets that are `in flight' at the end of a timestep are
stored on a stack, to be processed as part of the RT during the
subsequent timestep. The total number of photon packets is the sum of
the number of packets on the stack ($N_s$), and the number of new
photon packets generated during the timestep $N_p$. The path
of each photon packet is followed as it propagates through the gas,
during which time the packet may be scattered multiple times. The
random walk of the photon packet ceases when it is either absorbed or
leaves the computation domain (in which case the photon packet is
destroyed), or when its flight time ($l / c$) reaches $\Delta t$ (in
which case the photon packet is added to the stack). Once the random
walks of all the photons have been calculated, we may then use our MC
estimate of the absorption rate (equation~\ref{eq:new_absorption-rate}) to
update the gas energy density of each cell via equation
\ref{eq:ug-update}. One can immediately see that when the photon
flight time of a mean free path $\kappa \rho / c \gg \Delta t$ then
$N_s$ will dominate over $N_p$ as the calculation proceeds,
but eventually if radiative equilibrium is reached there will be an
approximately constant number of packets at each timestep.

It should be noted that, unlike the \cite{lucy_1999} radiative
equilibrium algorithm, energy is not implicitly conserved here, as
photon packets are created and destroyed during each timestep. The
quality of the energy conservation is controlled by the accuracy of
our estimator of the absorption rate (which in turn is dependent on
our MC statistics), and by our choice of $\Delta t$. Energy
conservation is addressed in section~\ref{sec:tests}.

The algorithm itself is described as a sequence of steps:
\begin{enumerate}
\item The current energy density of the gas ($u_g$) is used to compute
  the temperature distribution (equation \ref{eq:gas-energy-density})
  and thus the gas energy emission rate as given by equation
  \ref{eq:emissivity}. A probability distribution of a photon packet
  being emitted is calculated over all cells
\begin{equation}
p_i = \frac{\sum_1^i \dot{E}_i{V_i}}{\sum_1^{N_{\rm cells}}
  \dot{E}_i{V_i}}
\label{eq:gasprobdist}
\end{equation}
where $V_i$ is the volume of the $i$th cell.
\item The probability of a photon being produced in the gas is calculated
by
\begin{equation}
p_g = \frac{L_g}{L_* + L_g}.
\end{equation}
We take $\eta$ to be the uniform random deviate. If $\eta < p_g$ then
the photon packet is emitted within the gas, and its position is found
according to the probability distribution given by
equation~\ref{eq:gasprobdist}, otherwise the photon is produced by the
star.

\item If a photon is produced by the gas then the photon frequency is
  assigned according to the equation
\begin{equation}
\eta = \int_0^\nu j_\nu d\nu / \int_0^\infty j_\nu d\nu
\end{equation}
where $j_\nu = k_\nu B_\nu$. If the photon packet is stellar in origin
then its frequency comes from
\begin{equation}
\eta = \int_0^\nu B_\nu d\nu / \int_0^\infty B_\nu d\nu
\end{equation}
where $B_\nu$ is the Planckian at the stellar effective temperature
(although other stellar spectral energy distributions may be easily
incorporated).

\item A photon can propagate an optical depth 
\begin{equation}
\tau = (k_\nu + k_{\rm sca, \nu}) \rho \ell
\end{equation}
(where $k_{\rm sca, \nu}$ is the scattering coefficient at frequency
$\nu$) before it is either absorbed or scattered. The optical depth is
selected randomly:
\begin{equation}
\tau =-\ln (1-\eta)
\end{equation}
If $\tau$ is greater than the optical depth to the cell boundary then
the photon packet position is moved to the cell boundary and a new
optical depth is selected. If the time to travel optical depth $\tau$
is greater than $\Delta t$ then then photon packet is moved by
distance $c \Delta t$ and the photon information is placed on the
stack. Otherwise the photon position is updated and the type of
interaction is determined by the albedo
\begin{equation}
\eta \le \alpha = \frac{k_{\rm sca, \nu}}{k_\nu + k_{\rm sca, \nu}}.
\end{equation}
If the photon packet is scattered then the new direction vector
is chosen randomly from the appropriate phase function. If the photon
is absorbed then the appropriate path length is stored and the next
photon packet is selected.

\item Once all photons packets have been processed the new absorption
  rates are computed and the energy density of the gas is updated
  using equation \ref{eq:a-minus-e}. Subsequently the duration of the
  next timestep is determined (see below).

\end{enumerate}

\subsection{Timescales}
\label{section:timescales}
Cells in which the emission rate exceeds the absorption rate are
cooling, and the timescale for cooling is calculated as
\begin{equation}
t_{\rm cool} = \frac{u_g}{\dot{E}-\dot{A}}
\end{equation}
Conversely, if the energy absorption rate exceeds the emission rate
for a given cell then its temperature is increasing. We may then
calculate an approximate radiative equilibrium timescale by first computing the
equilibrium energy density $u_e$ (using equation \ref{eq:rad-eq-time})
and then finding
\begin{equation}
t_{\rm eq} = \frac{u_e - u_g}{\dot{A} - \dot{E}}
\end{equation}
Cells for which $t_{\rm eq} \le \Delta t$ have their temperatures set
to the radiative equilibrium temperature. Note that this formalism assumes that
$\dot{A}$ and $\dot{E}$ are constant over $\Delta T$, and care must be taken to ensure that one does not erroneously set cells to their equilibrium temperature due to a poorly chosen $\Delta T$.

\section{Tests}
\label{sec:tests}
I have implemented a number of test cases in order to examine the
efficacy of the new algorithm.

\subsection{Radiative equilibrium}
\label{sec:test1}

\begin{figure}
  \includegraphics[width=85mm]{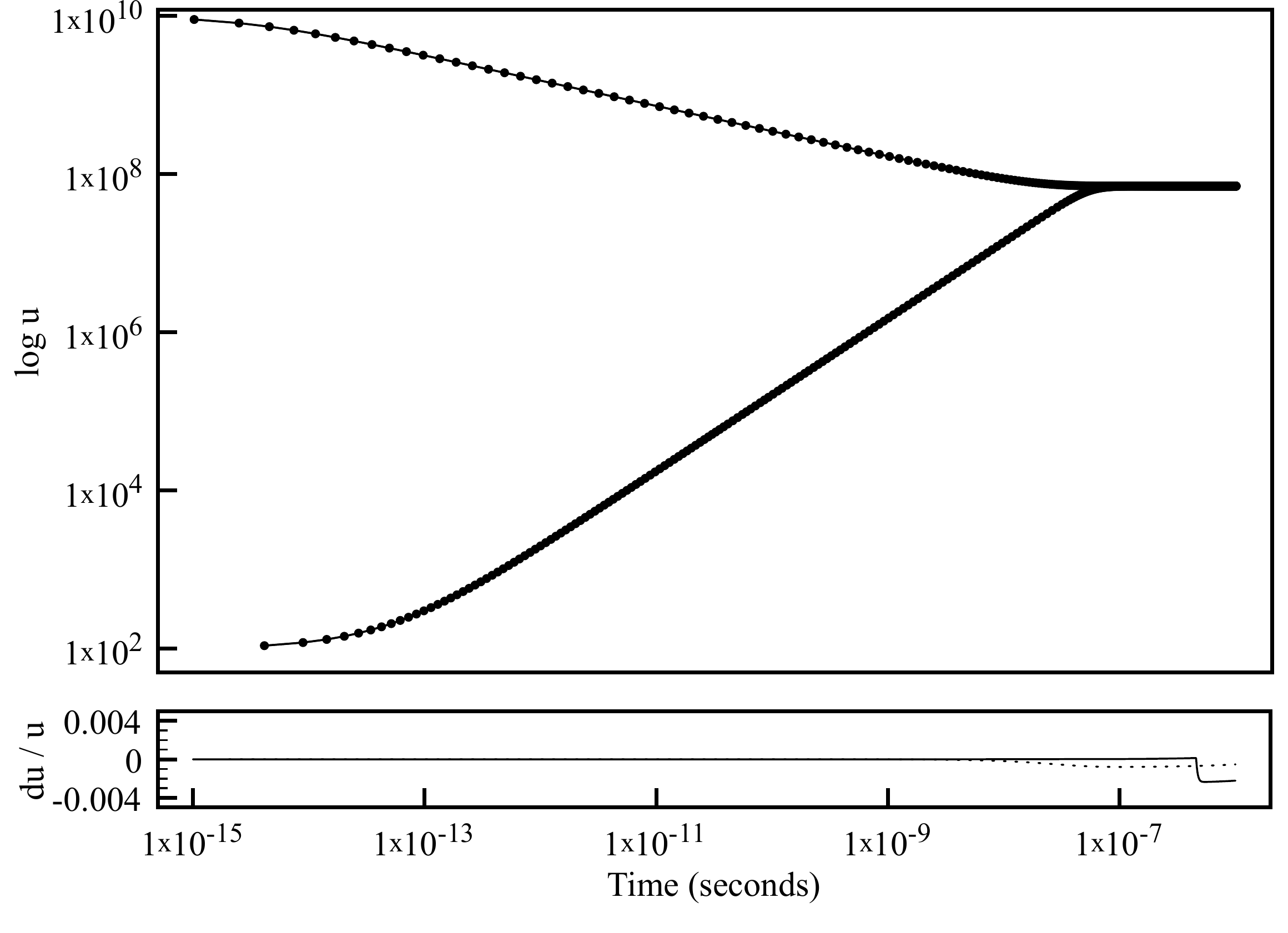}
  \caption{The results of the first radiative equilibrium test case
    described in Section~\ref{sec:test1}. In the upper panel the Monte-Carlo results are
    shown as dots, while the analytical solution found from
    equation~\ref{eq:analytical1} is displayed as a solid line. The bottom panel shows the fractional difference between the analytical and numerical solutions for the energy density. }
  \label{fig:fig1}
\end{figure}

If we immerse an absorbing gas in a radiation field of much higher
energy density the gas will eventually come into 
equilibrium with the radiation. Since $u_g \ll u_r$ we can assume $u_r$
is a constant and the evolution of $u_g$ can then be found from the
ordinary differential equation
\begin{equation}
\frac{du_g}{dt} = c \kappa u_r - 4 \pi \kappa B(u_g).
\label{eq:analytical1}
\end{equation}
For this test we adopted the parameters given by \cite{turner_2001}:
$\rho=10^{-7}$ g\,cm$^{-3}$, $\kappa = 4 \times 10^{-8}$ cm$^{-1}$,
$\mu = 0.6$, $\gamma = 5/3$ and $u_r = 10^{12}$ erg\,cm$^{-3}$. We ran
  two separate cases, one with $u_g$ intially well below the
  equilibrium value ($u_g=10^2$ erg\,cm$^{-3}$) and one with it well
  above ($u_g=10^{10}$ erg\,cm$^{-3}$). The results are plotted in
  Figure~\ref{fig:fig1} and the agreement with the analytical solution
  is excellent.

\begin{figure}
  \includegraphics[width=85mm]{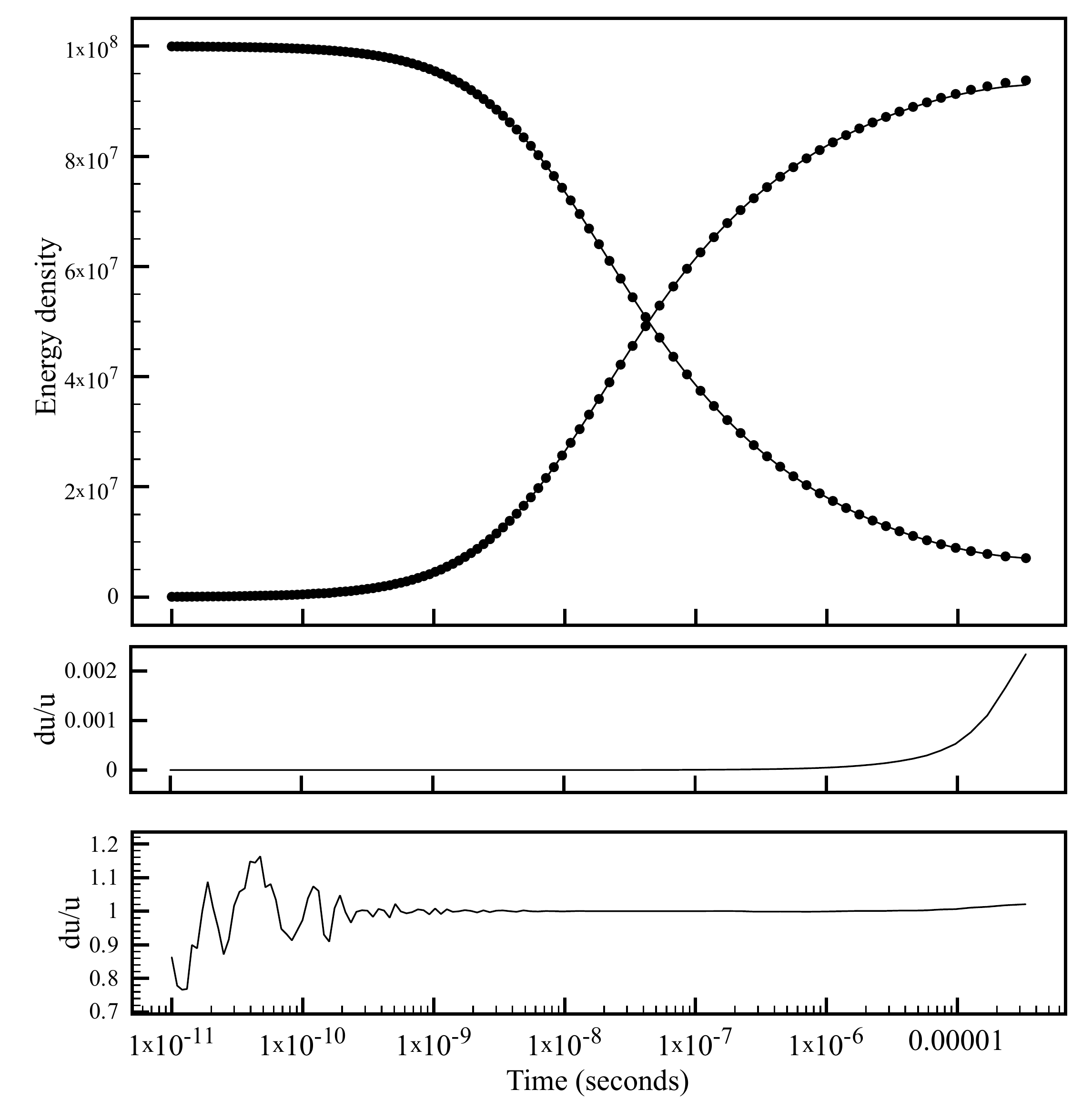}
  \caption{The results of the second radiative equilibrium test case
    described in Section~\ref{sec:test1}. In the upper panel the Monte-Carlo results are
    shown as solid lines, while the solutions found from numerical
    integration of equations~\ref{eq:e-minus-a} and~\ref{eq:a-minus-e}
    are displayed as dots. The middle and lower panels shows the fractional difference between the energy densities from the analytical and Monte Carlo solutions for equations~\ref{eq:e-minus-a} and~\ref{eq:a-minus-e} respectively.}
  \label{fig:test2_fig}
\end{figure}

The second test involved the same physical parameters for the box
detailed above, but instead of filling the box with a high density
radiation field we instead set $u_r = 0$ and $u_g = 10^8$ everywhere.
As the RT is followed the material in the box will cool and emit
radiation. Since we have reflective boundary conditions for the photon
packets the gas and radiation field will eventually settle into
radiative equilibrium. We set $N_p$ to 1, which means that the gas
emits one photon per timestep. In Figure~\ref{fig:test2_fig} we plot
the evolution of the radiation and gas energy densities with time, and
these show good agreement with a simple numerical integration of
equations~\ref{eq:e-minus-a} and \ref{eq:a-minus-e}. We find that the
total energy is conserved to better than 0.5\% throughout the duration
of the MC calculation.

The above tests suggests that the physics of matter-radiation
interaction is adequately captured by the new algorithm, and we now
progress to tests of the transport of radiation.

\subsection{Diffusion limit}

In the optically-thick limit the radiation transport occurs according
to the diffusion equation
\begin{equation}
\frac{du_r}{dt} = - D \frac{d^2u_r}{dx^2}
\label{diff_eq}
\end{equation}
where $D$ is the diffusion coefficient, given by
\begin{equation}
D = \frac{c}{\kappa}.
\end{equation}
Here we test the Monte-Carlo algorithm on a heat kernel problem, in
which energy is deposited at a point at $t=0$\,s and is allowed to diffuse
through the medium. The analytical solution to the diffusion equation
for this scenario is a gaussian
\begin{equation}
u(x,t)  = \frac{1}{\sqrt{4 \pi D t}} \exp \left( - \frac{x^2}{4 D t} \right).
\end{equation}

We assume a 1\,cm one-dimensional box, divided into 101
evenly-spaced bins, with reflective boundary conditions for the
photons (i.e. adiabatic). We adopt $\kappa = 10^{13}$\,cm$^{-1}$ and deposit
$10^{10}$ of energy in photons into the central bin at $t=0$\,s. (The
radiation field immediately comes into radiative equilibrium with the
material in the box, but $u_r \gg u_e$).  We follow the radiation
transport using the Monte-Carlo algorithm, and also by solving
equation~\ref{diff_eq} using the Crank-Nicolson method.

\begin{figure}
  \includegraphics[width=85mm]{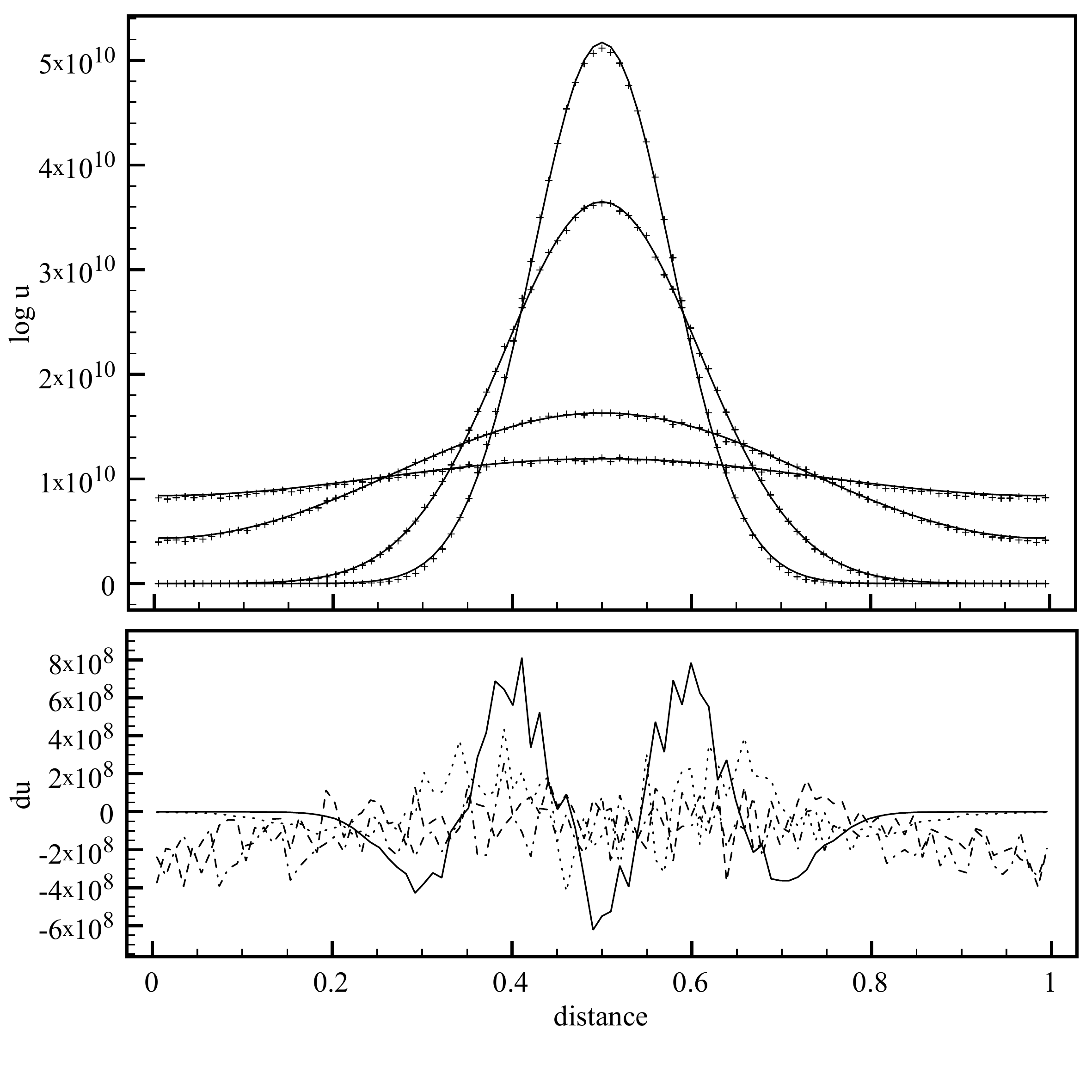}
  \caption{Heat kernel test for the pure scattering case. In the upper panel the solution
    to the diffusion equation is plotted (solid lines) for times $t =
    10^{-11}$, $2 \times 10^{-11}$, $10^{-10}$, and $2 \times
    10^{-10}$\,s, while the radiation energy densities for the same times found
    using the MC algorithm are plotted as crosses. The lower panel shows the absolute differences between the analytical and MC-based energy densities at the same timesteps (solid, dotted, dashed, and dot-dashed lines respectively).}
  \label{fig:test4_fig}
\end{figure}

We initially assume pure scattering ($\alpha = 1$), in which case the
MC algorithm will always perfectly preserve energy since no photons
are created or destroyed and matter/radiation energy transfer terms
are by definition zero. This scenario tests the random walk and photon
flight time limit section of the MC algorithm, and  we find good agreement
between the diffusion approximation and the MC algorithm for this
test case (Figure~\ref{fig:test4_fig}).

\begin{figure}
  \includegraphics[width=85mm]{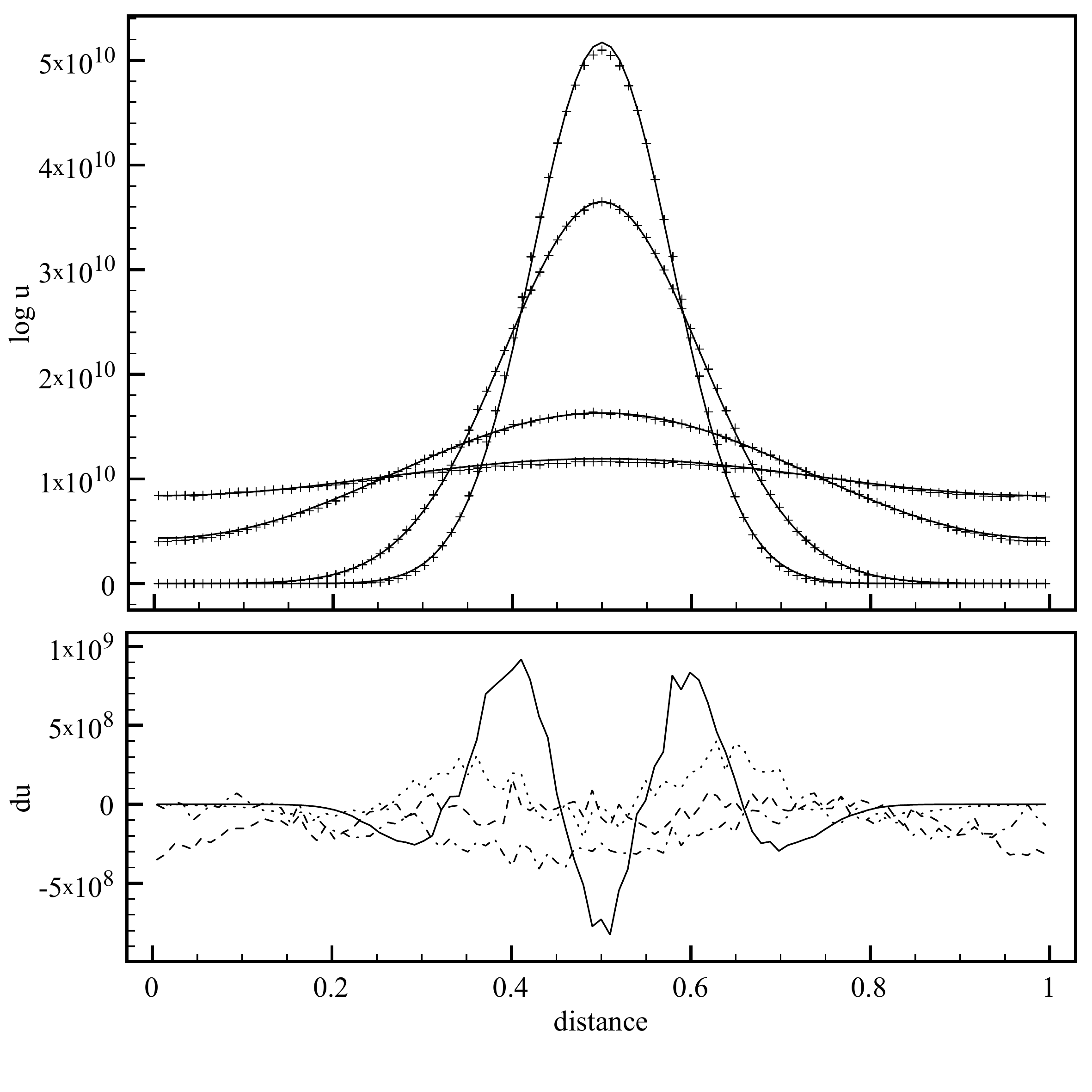}
  \caption{Heat kernel test for the pure absorption case. In the upper panel the solution
    to the diffusion equation is plotted (solid lines) for times $t =
    10^{-11}$, $2 \times 10^{-11}$, $10^{-10}$, and $2 \times
    10^{-10}$\,s, while the radiation energy densities for the same times found
    using the MC algorithm are plotted as crosses. The lower panel shows the absolute differences between the analytical and MC-based energy densities at the same timesteps (solid, dotted, dashed, and dot-dashed lines respectively).}
  \label{fig:test3_fig}
\end{figure}

The purely absorptive case ($\alpha=0$) is a much more challenging
proposition. Both the emissivity and absorption rates are very high,
and Monte-Carlo sampling errors can potentially lead to deviations
from energy conservation. However we find good agreement
(Figure~\ref{fig:test3_fig}) with the diffusion approximation, and
energy is conserved to within 2~percent even after more than 4000
time steps.

\subsection{Time varying source}

It appears that the diffusion limit is well-modelled using the MC
algorithm, but such conditions are naturally more efficiently treated
using the diffusion approximation. However many astrophysical problems
involve transport through substantial regions of optically thin
material, and here we conduct a test of the MC algorithm that
incorporates both optically thin and opaque material. We also include
a time varying source of photons which allows us both to test how well
the algorithm retains the coherence of the radiation field but also
how well it captures the heating and cooling of material.

We adopt a 1\,cm box with a density of $2 \times
10^{-5}$\,g\,cm$^{-3}$ and
\begin{equation}
\kappa(x) = \left\{ 
\begin{array}{l l}
  0 & \quad \mbox{if $x<0.5$}\\
  10^6 & \quad \mbox{if $x \ge 0.5$}\\ 
\end{array} \right.
\end{equation}
Photons are injected into the left-hand boundary of the box from a
source of luminosity that varies sinusoidally between zero and
$10^{20}$\,erg\,s$^{-1}$ on a period of $10^{-11}$\,s. Photons
propogate through the optically-thin part of the box until they
encounter the optically thick-material and are absorbed, thus heating
the right hand half of the box. This heated material emits radiation
in an attempt to come into thermal balance; since the optical depth is
smaller in the $-x$ direction these photons are preferentially emitted
in that direction (accompanied by a transport by diffusion in the $+x$
direction). Outflow boundary conditions are imposed either end of
the box.

\begin{figure}
  \includegraphics[width=83mm]{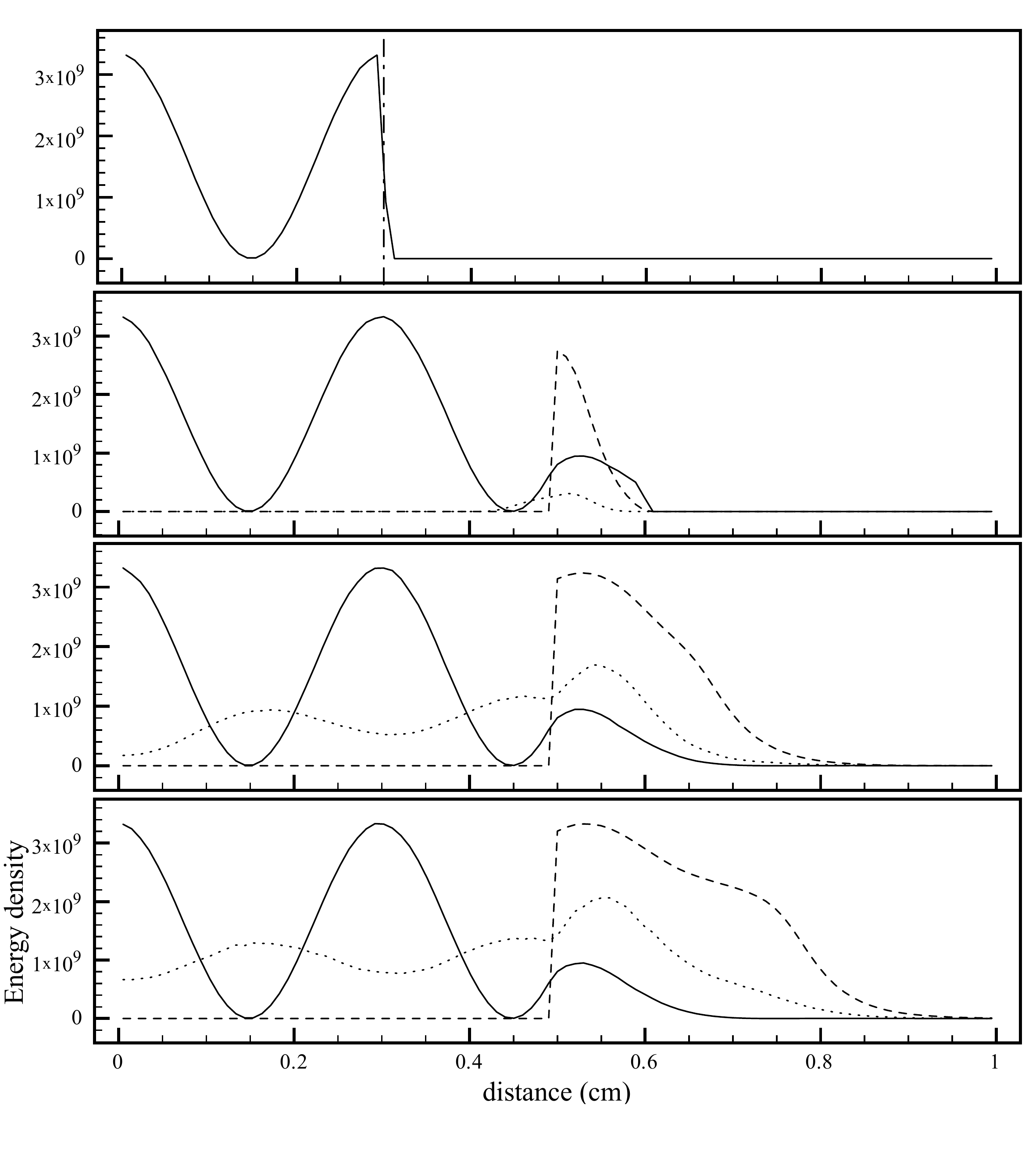}
  \caption{An optically thin/thick radiation propogation test with a
    varying source. The panels (top to bottom) are snapshots at $t =
    10^{-11}$, $2 \times 10^{-11}$, $4 \times 10^{-11}$ and $6 \times
    10^{-11}$\,s. Plotted are the radiation energy density of the
    source photons (solid line), the thermal energy density of the
    material (dashed line), and the radiation energy density of
    photons emitted by the material (dotted line). The expected
    maximum extent of the radiation field from the source at $t =
    10^{-11}$\,s is plotted as a dot-dashed line in the upper panel.}
  \label{fig:test5_fig}
\end{figure}

Several timesteps from this test calculation are presented in
Figure~\ref{fig:test5_fig}. Initially the sinusoidally varying
radiation field can be seen to propagate at speed $c$ through the
left-hand half of the box. The optically thick material is heated and
re-emits radiation which can then be seen propagating to the left
with a sinusoidally varying flux (the amplitude of the variability is
lower than that of the impinging radiation field, never reaching zero,
due to thermal capacity of the optically thick material). The
diffusion of radiation through the optically thick material is also
apparent. 

\begin{figure}
  \includegraphics[width=83mm]{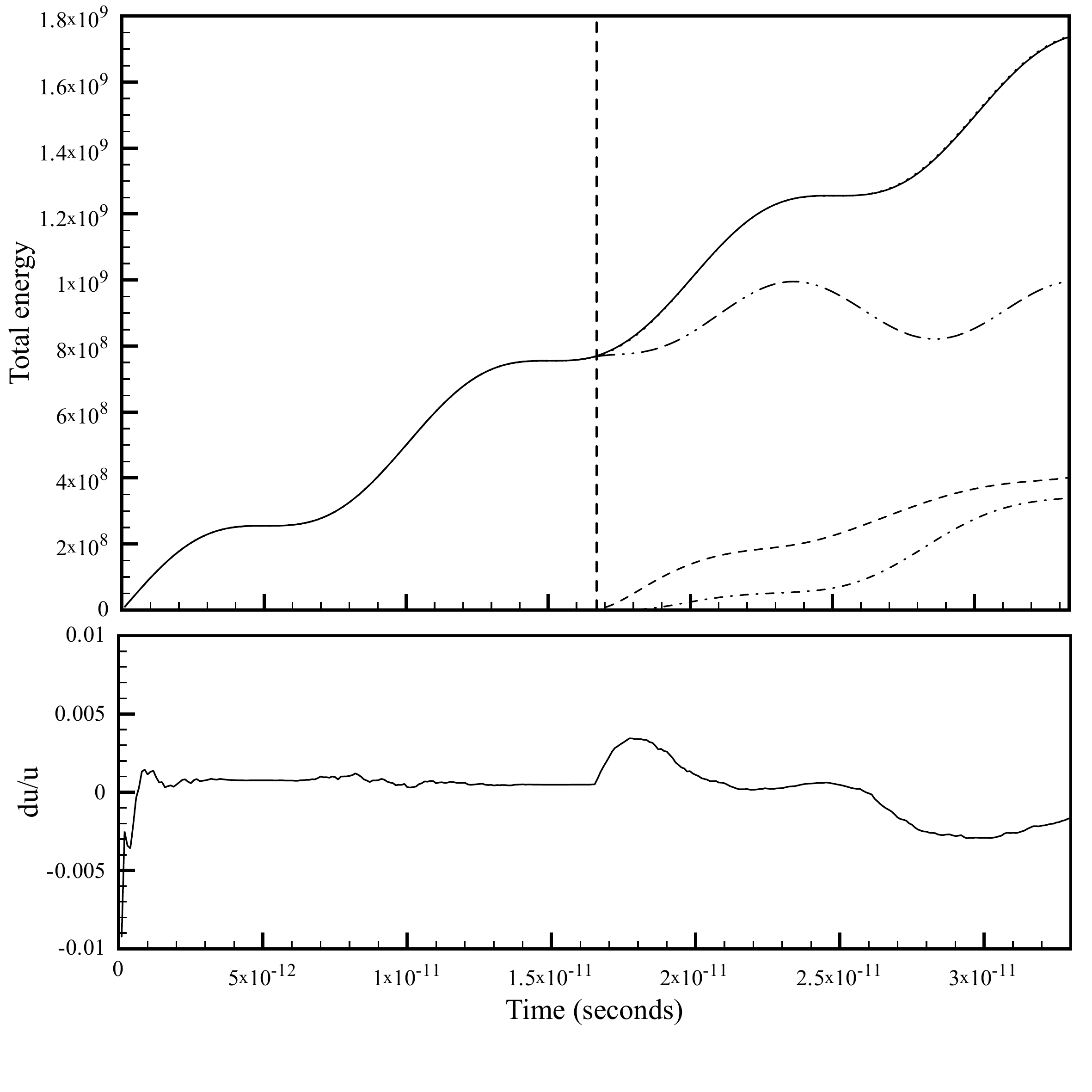}
  \caption{Total energies for the the optically thin/thick radiation
    propogation test with a varying source. In the upper panel the expected total energy
    in the box is plotted as a solid line, along with  the total energy
    calculated from the MC algorithm (dotted line, very close to the
    solid line),  the energy in the
    radiation field (dot-dot-dash line), the thermal energy (dashed
    line), and the radiation density emitted from the heated material
    (dot-dash line). The vertical dashed line indicates the expected
    time that the source radiation field enters the optically thick
    material ($1.67 \times 10^{-11}$\,s). The lower panel shows the fraction difference between the expected total energy in the box and that derived from the MC-method.}
  \label{fig:test5_fig2}
\end{figure}

We have checked the energy conservation of the algorithm as the source
radiation is transported, absorbed and remitted. In
Figure~\ref{fig:test5_fig2} shows the integrated thermal and radiation
energy presented in the box at each timestep (prior to any radiation 
escaping the box at the boundaries). The total energy is conserved to
better than 1\% at all timesteps.

\subsection{Implementation}

The above 1-d, grey tests indicate that the algorithm is fundamentally
sound and can reproduce analytical results. We therefore implemented
version of the algorithm as a module within the {\sc torus} radiative
transfer code \citep{harries_2000, kurosawa_2004, pinte_2009}. The code is written
in Fortran~90 and follows the radiative transfer on an adaptive mesh stored as a quad-tree (2-d) or oct-tree (3-d).

Since each Monte-Carlo photon packet is essentially an independent event the code is straightforwardly parallelized under the Message Passing Interface (MPI). Each thread holds a copy of the grid in memory, and the work of the photon loop is distributed over all the MPI threads. At the end of the photon packet loop the sums in equations~\ref{eq:energy-summation} and \ref{eq:new_absorption-rate} are made over all threads, and thus estimates for the photon energy density and absorption rate are found for each cell in the grid, and finally the results are distributed back to all threads. The communication overhead for the last step is minimal compared to the calculations for the photon loop, and the calculation is thus CPU rather than thread-communication limited. 

Example CPU times are susceptible to rapid erosion in usefulness due to Moore's law, but for the benefit of contemporary readers the science grade calculation described in section~\ref{application_sec} was run on 16 theads on the University of Exeter's SGI Altix ICE 8200 supercomputer and each of the later timesteps (when the photon stack size had maximized) took approximately 4 minutes, of which 20 seconds were inter-thread communication. It was found that the calculation scaled almost perfectly up to 64 threads (the maximum that we tested).

\subsection{Benchmark protostellar disc}

As an
initial test we used the 2-d RT benchmark disc described by
\cite{pascucci_2004}. By following the radiation-transport over a
sufficiently long duration the disc (initially at $T=0$\,K) will come
into radiative equilibrium with the central star. This represents a
rigorous test of the new algorithm since the circumstellar material
contains both optically thin and optically thick regions and thus
simultaneously incorporates both free-streaming radiation and
diffusing photons. Furthermore the polychromatic nature of the
algorithm is tested with excess luminosity from the photosphere at
short wavelengths heating the disc and being re-emitted in the mid-IR
and sub-mm regimes.

\begin{figure*}
\begin{center}$
\begin{array}{cc}
  \includegraphics[width=88mm]{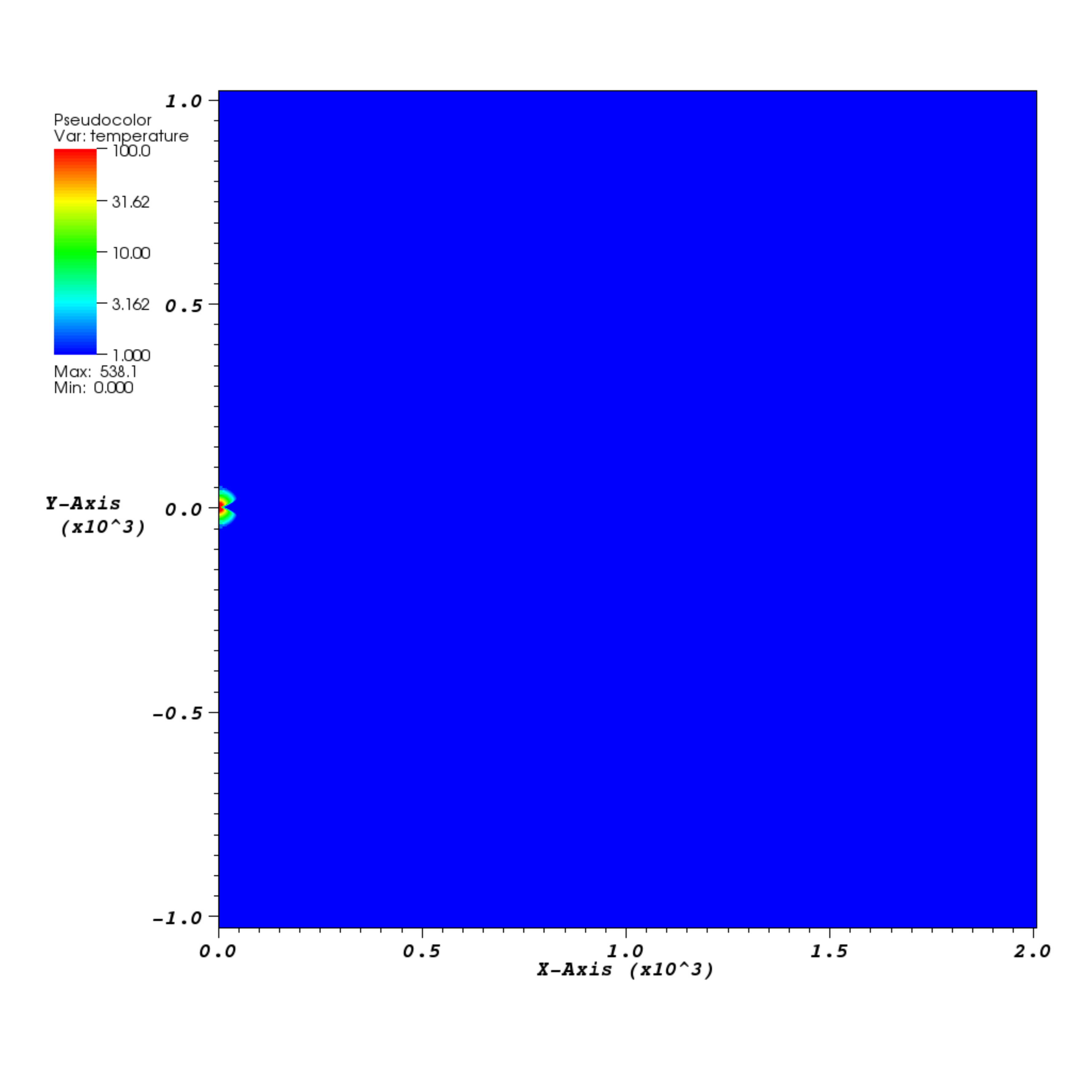} &
  \includegraphics[width=88mm]{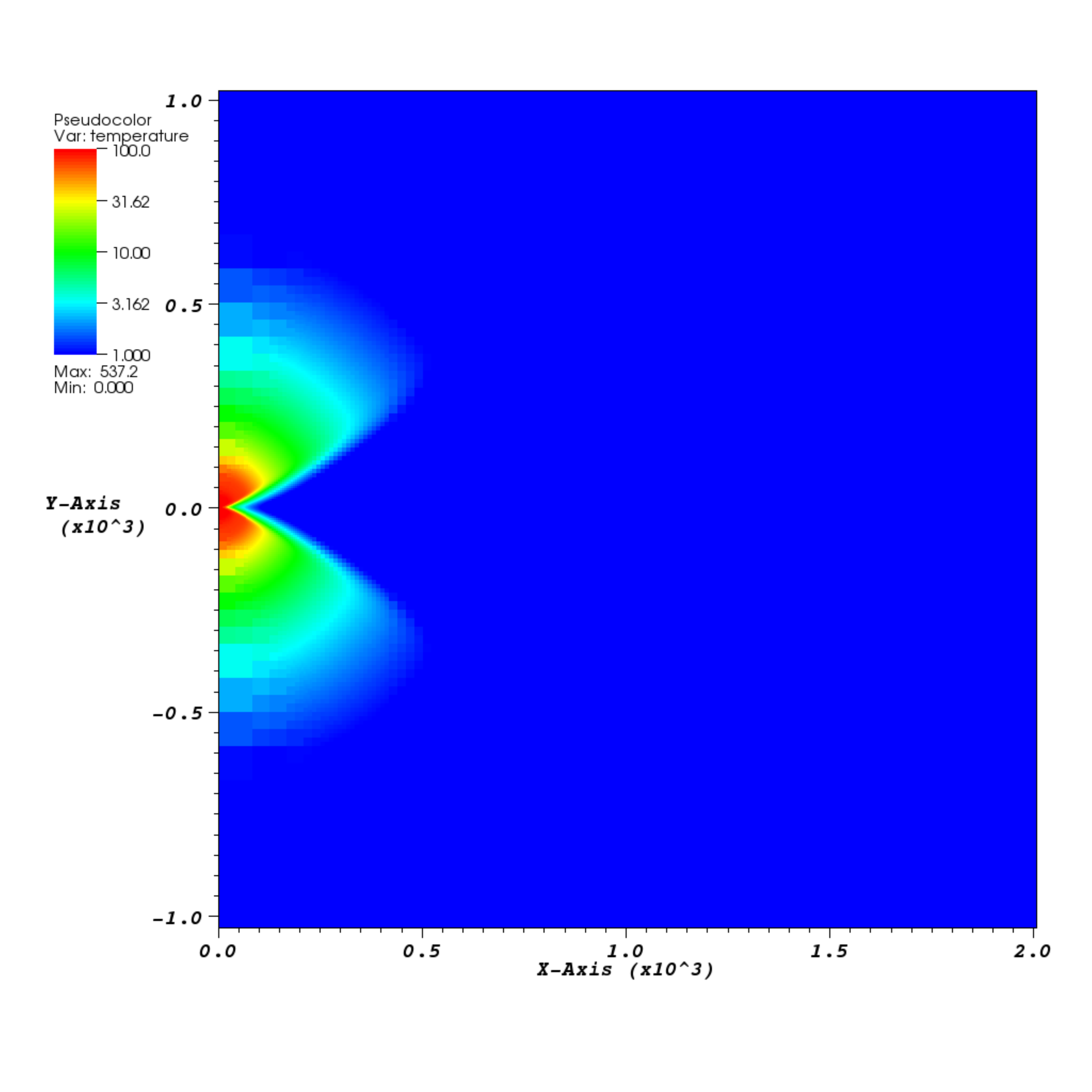} \\
  \includegraphics[width=88mm]{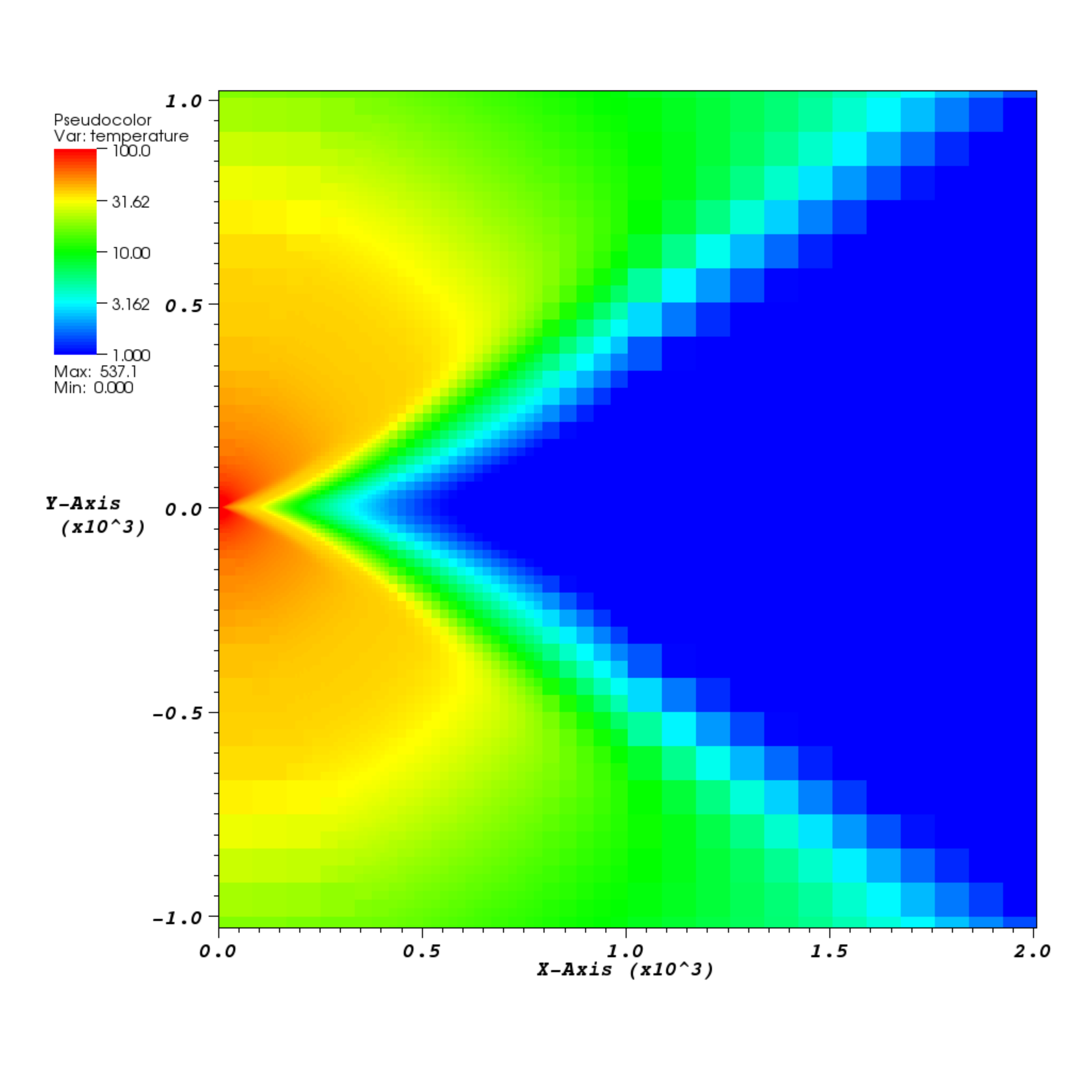} &
  \includegraphics[width=88mm]{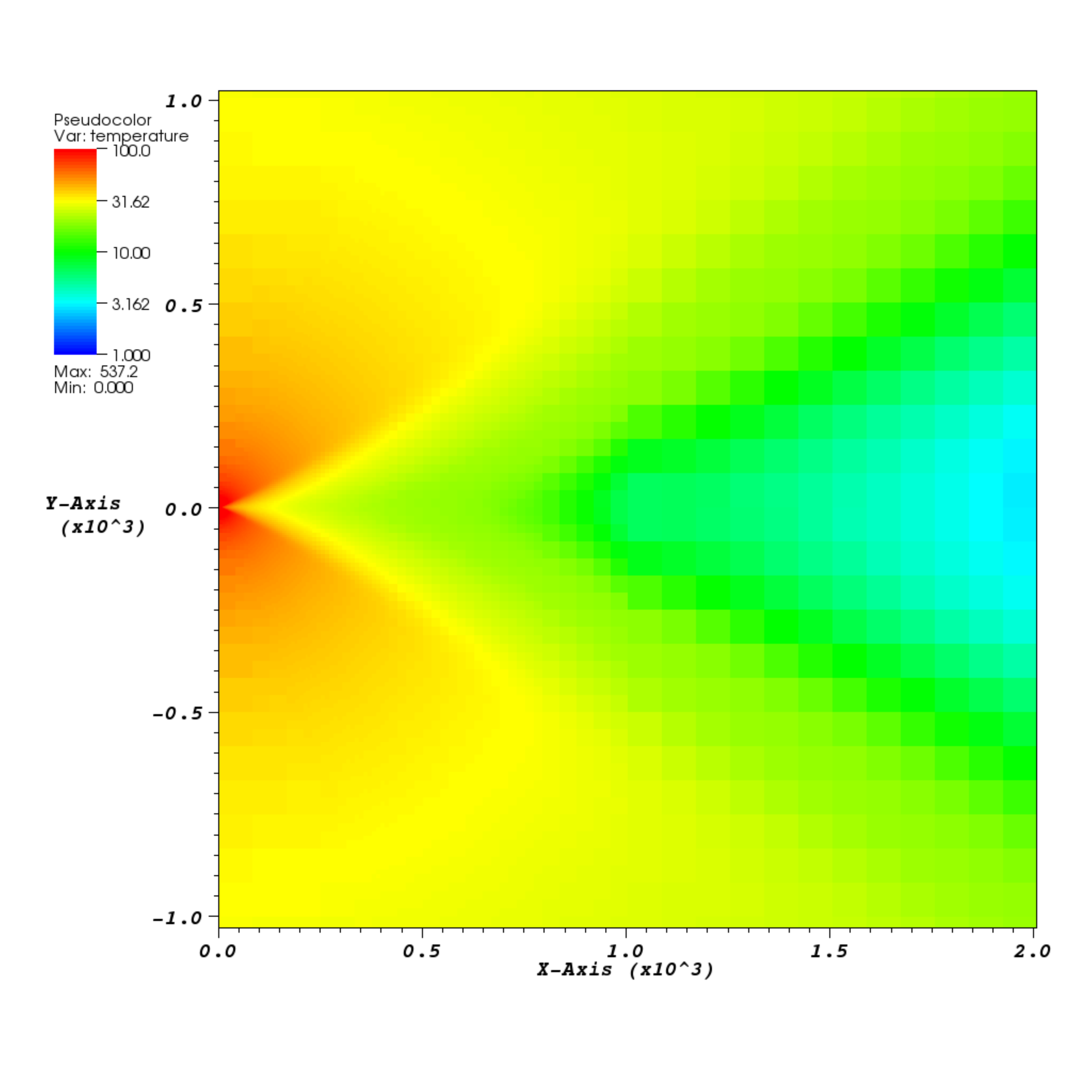}
\end{array}$
\end{center}
  \caption{The Pascucci benchmark disc heated from $T=0$\,K to thermal equilibrium using
the MC algorithm. The temperature of the disc is plotted as a
logarithmic colour-scale scaled  between 1 and 100\,K. Distances are
in AU. Four snapshots are shown: $t=3.1 \times 10^4$\,s (top left panel), $t=1.02
\times 10^6$\,s (top right), $t=3.3\times10^7$\,s (bottom left), and
$t=1.05 \times 10^{9}$\,s (bottom right).}
\label{fig:pascucci_colour}
\end{figure*}

We adopted an initial timestep of 100\,s, and used $10^6$ photons per
timestep. In Figure~\ref{fig:pascucci_colour} the gradual heating of
the disc is shown. The rarified regions above and below the disc
quickly reach thermal equilibrium, while the disc midplane, which has
a characteristic radial optical depth of $\sim 100$ at 5500\,\AA,
approaches equilibrium much more slowly. After $\sim 3 \times
10^{10}$\,s the entire disc has reached a steady state.

\begin{figure}
  \includegraphics[width=85mm]{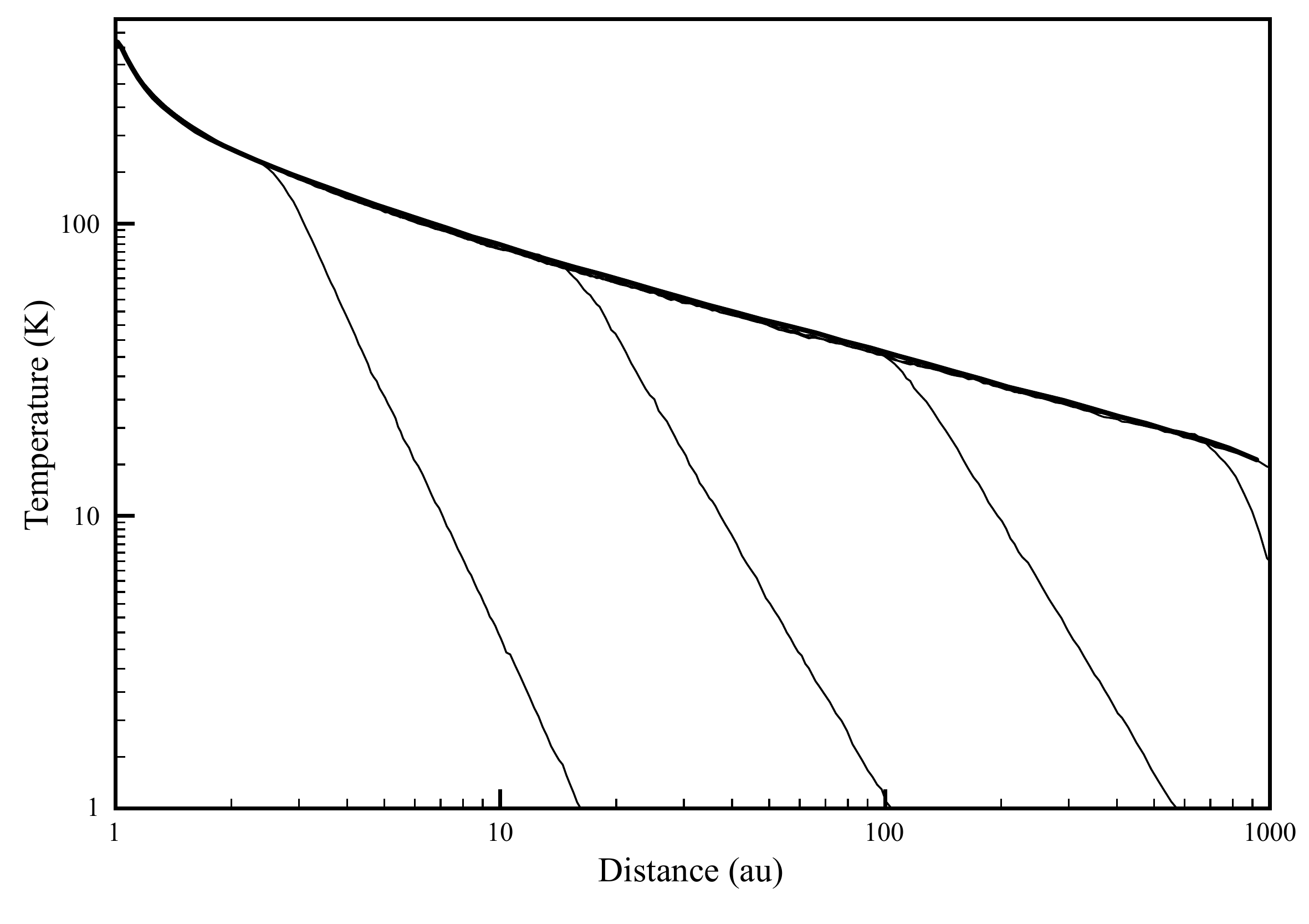}
  \caption{The disc midplane temperature for the Pascucci benchmark
    disc. The heavy line shows the benchmark temperature distribution,
    while the other lines show (from the left) snapshots at $t=3.1
    \times 10^4$\,s, $t=1.02 \times 10^6$\,s, $t=3.3 \times 10^7$\,s,
    $t=1.05\times10^9$\,s and $t=3.3 \times 10^{10}$\,s.}
  \label{fig:pascucci_radial}
\end{figure}

A quantitative comparison with the benchmark solution is shown in
Figure~\ref{fig:pascucci_radial}. Good agreement with the published
benchmark midplane temperature distribution is found after $\sim 10^{10}$\,s.

\section{Application: protostellar disc variability}
\label{application_sec}

The accretion rate onto protostars is inherently variable on a wide
variety of timescales from hours to years
(e.g. \citealt{bouvier_2007,nguyen_2009}). This variable accretion
flux (emitted primarly at wavelengths less than 4000\AA) is scattered and reprocessed in
the disc to be emitted in the near- and mid-IR. There will be an
intrinsic delay between the optical variability and the disc's IR
response, in part due to the light-travel time from the star to the
disc, but also due to thermal lag within the disc itself. Since the
temperature in the disc decreases radially, one expects the time lag
(with repect to the optical flux variability) to increase with
wavelength.  Hence by obtaining simultaneous timeseries photometry in
the optical and IR one may attempt to map the disc emission using the
wavelength-dependent lag, in much the same way as reverberation
mapping is used to map the broad-line region around AGN from the
emission line/UV continuum variability
(e.g. \citealt{denney_2009}). Of course there are a host of other
factors that might complicate this rather naive
interpretation of the variability, such as changes to the disc
structure, and rotational modulation of either an azimuthally
structured inner edge (warps) or an anisotropic radiation field (hot
spots), but it is nonetheless interesting to examine the expected
timescales from a simple thermal response model.

\subsection{Model parameters}

We adopt a central protostar of radius $R_* = 2$\,R$_\odot$ and a
blackbody photospheric flux distribution at $T_{\rm eff} =
4000$\,K. The recent study by \cite{nguyen_2009} showed that the
accretion rate onto a typical protostar varies by a factor of $\sim 2$
on a timescale of days to weeks. We assume the star is accreting at a
rate that varies sinusoidally between $5 \times
10^{-8}$\,M$_\odot$\,yr$^{-1}$ and $1 \times
10^{-7}$\,M$_\odot$\,yr$^{-1}$ with a period of 1\,h.   This period is comparable with the flushing timescale of the magnetosphere and line profile variability on such timescales has been observed  e.g. \cite{smith_1999, kurosawa_2005}. Such rapid variability corresponds
to a change on a much shorter timescale than the canonical rotation
period, making it easier observationally to distinguish disc
reprocessing effects from rotationally-modulated disc-structure
effects.

 In order to mimic the additional flux associated with the
accretion we add the accretion luminosity as a blackblody with a
characteristic temperature found by assuming that the accretion power
is emitted from an area equivalent to 5\% of the stellar photosphere.

A simple flared structure is adopted for the disc, viz:
\begin{equation}
\rho(r,z) = \rho_0 r^{-\alpha} \exp \left(  - \frac{1}{2} \frac{z^2}
  {h(r)^2} \right)
\end{equation}
where
\begin{equation}
h = h_0 (r/r_0)^\beta
\end{equation}
with $\alpha = 2.25$ and $\beta = 1.25$. The disc scale height is set
to $h_0=125$\,au at $r_0=100$\,au. In order to simulate the truncation of
the inner disc by the magnetosphere we assume an inner hole of
10\,R$_*$, while the outer radius is 300\,au. We fix $\rho_0$ by
assuming a disc mass of 0.01\,M$_\odot$. The disc is assumed to
contain the dust size distribution and chemisty described by
\cite{kurosawa_2004}, and for the purposes of this test we assume
isotropic scattering.

\subsection{Method}

The disc is initially brought into radiative equilibrium with the
central object (assuming a constant accretion rate of $5 \times
10^{-8}$\,M$_\odot$\,yr$^{-1}$) by using a time-independent
algorithm. The time-dependent method is then turned on (which defines
$t=0$\,s) and the RT is followed for ten periods at a timestep of 18
seconds. This short timestep allows us to adequately resolve the
shortest timescales that are of interest, which is basically the
light-crossing time from the central object to the inner disc (47\,s).

Since the photon packet estimate of the radiation field is only
defined within a sphere of radius $ct$ around the central object it is
only cells within this radius whose thermal properties are changed at
each timestep, with the rest of the disc held in thermal
equilibrium. Free-streaming photon packets that have a negligble
probability of interacting with the disc (found by integrating the
optical depth along the packet's path) are deleted from the stack at
the end of each timestep. This has the advantage of significantly
reducing the memory requirement of the photon stack.

The shape of the cross correlation function should be a function of
the system's viewing angle (inclination) as different parts of the
disc have different star-disc-observer path lengths. In the following
we adopt an inclination of 60$^\circ$ when calculating the SEDs.

Initially a set of photon packets were generated from the volume of
grid outside the $ct$ sphere defined by the length of the
timeseries. Since this volume is at constant temperature it was
possible to calculate this contribution to the SED in a time-independent
manner. Subsequently at each timestep $10^6$ new photon packets
(photospheric or thermal disc) were generated, and these photons were
tagged by their generation time. A `peel-off' technique was then
employed: The light travel time and optical depths to the observer
were calculated, along with the probability that the photon packet was
emitted towards the observer direction. The photon packets were then
binned in the observer's frame, both spectrally and temporally. The
new photon packets, plus the photon packet stack, where then followed
for a single timestep, with packet peel-off and binning occurring at
each scattering event.

\subsection{Results}

\begin{figure}
  \includegraphics[width=80mm]{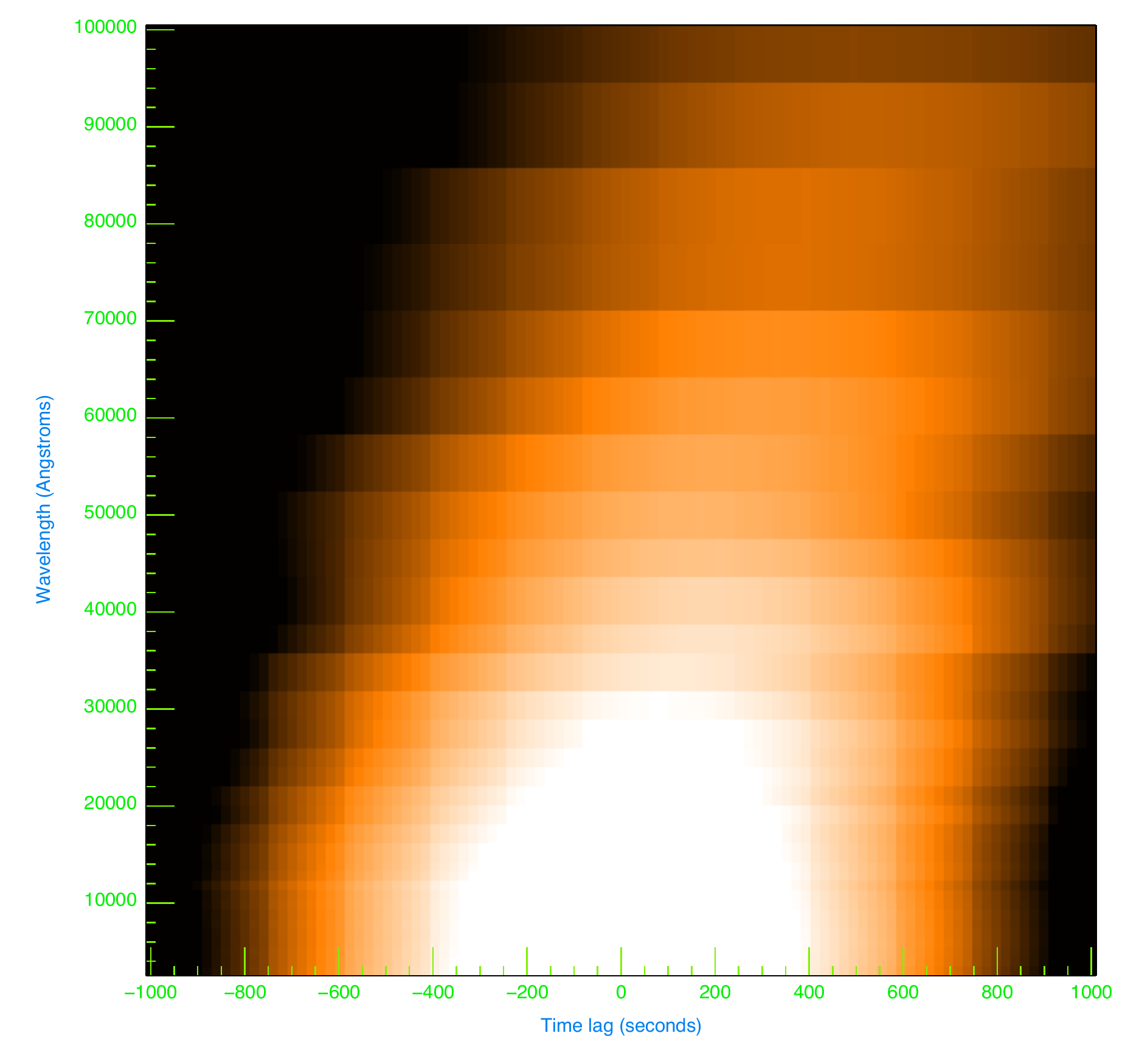}
  \caption{A cross-correlation image of the monochromatic disc
    lightcurve at 3000\,\AA\ against wavelengths from 3000\,\AA\ to
    10\,$\mu$m. The image is a linear greyscale of cross-correlation
    function scaled between 0.8 (white) and 1 (black). The bottom row
    of the image is the autocorrelation function of the 3000\AA\ lightcurve.}
  \label{fig:lag_greyscale}
\end{figure}

We have computed the cross-correlation function (CCF) of monochromatic
time-series from 3000\AA\ to 10\,$\mu$m against a fiducial time-series
at 3000\AA\ (see Figure~\ref{fig:lag_greyscale}). It is immediately
apparent that the lag increases with wavelength due to a combination
of light travel time to the disc, and the thermal lag as the increased
radiation field heats the local disc material. The maximum value of
the cross correlation function decreases at longer wavelengths as the
thermal timescale of the disc begins to dominate over the variability
timescale.

\begin{figure}
  \includegraphics[width=85mm]{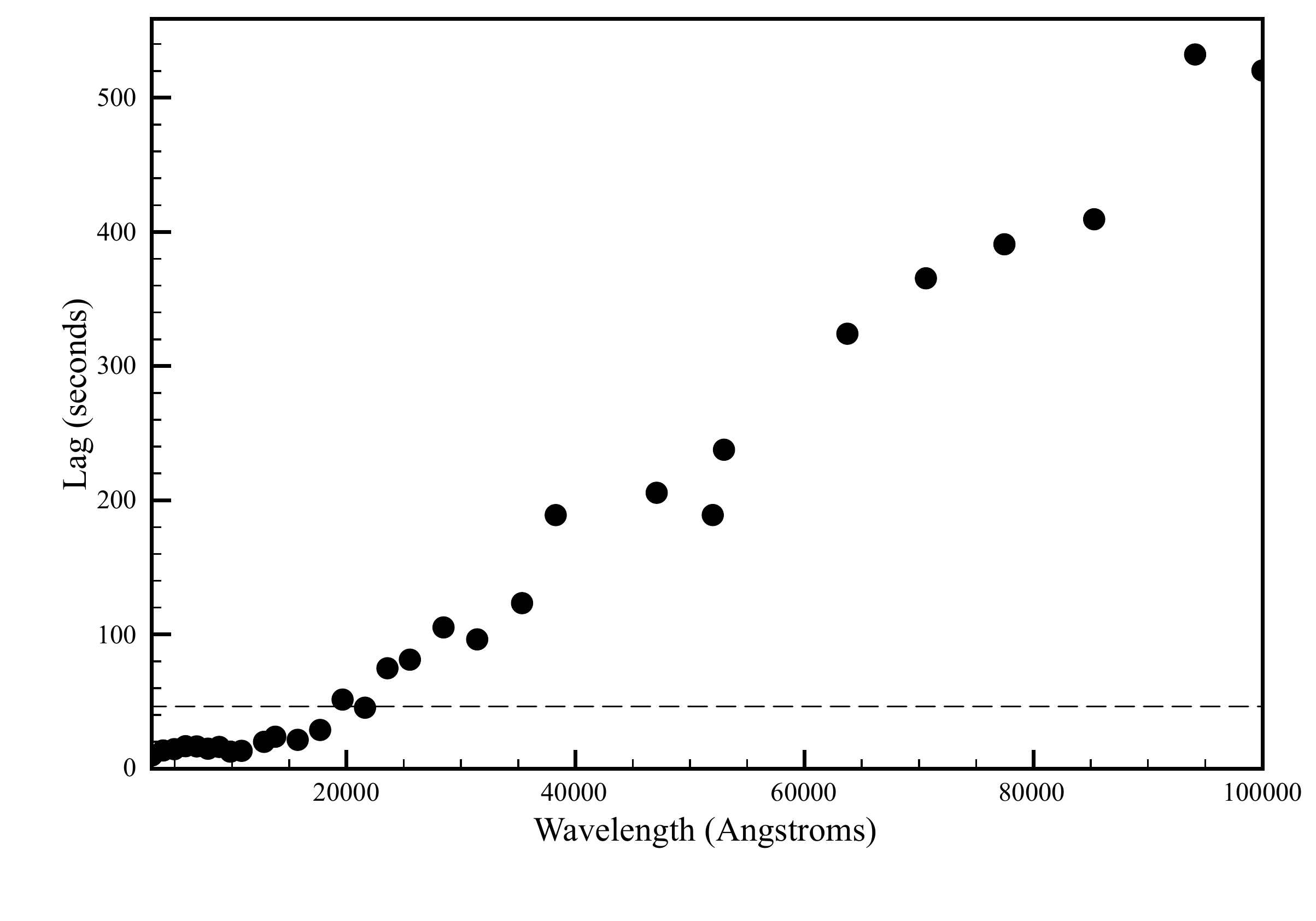}
  \caption{The time lag with respect to the 3000\AA\ lightcurve
    plotted as a function of wavelength (filled circles). The light
    travel time between the stellar centre and the inner disc edge is
    plotted as a dashed-line.}
  \label{fig:lag_graph}
\end{figure}

 I have quantified the lag by fitting a Gaussian to each CCF peak as
 a function of wavelength. The central positions of these Gaussians
 are plotted in Figure~\ref{fig:lag_graph}. The lags at blue
 wavelengths ($\ll 2$\,$\mu$m) are short, in fact less then the light travel
 time between the star and the disc, this is principally because there
 is significant direct photospheric emission at these wavelengths
 (with zero lag) in addition to the scattered and reprocessed
 components. Redwards of $\sim 2$\,$\mu$m the flux is dominated by
 thermal emission from the disc, and we see that the lag increases
 monotonically (within the errors on the CCF Gaussian  fits) with
 wavelength (although the strength of the correlation is decreasing). 

Observing this correlated phenomenon would be a challenging
proposition, necessitating high cadence temporal sampling in the
optical and near-IR simultaneously. Optimally one would select
ground-based $UBV$ photometry combined with {\em Spitzer} IRAC
photometry. Although it may be possible to conduct the IR observations
into the mid-IR it is clear that the correlation is dropping rapidly
at these wavelengths, and it is not at all clear the necessary
temporal sampling could be achieved using ground-based facilities.
It may be that systems with a larger inner-hole, such as Herbig AeBe stars \citep{monnier_2005} would be superior targets to Classical T~Tauri stars, allowing light-travel time effects to remain dominant at longer timescales.

Of course in reality the underlying accretion luminosity is unlikely
to vary periodically, but more stochastic variations could still be
investigated. It may be possible to distinguish between disc thermal
responses and scattering by observing polarized light, which will be
dominated by the scattered light contribution. 

\section{Conclusions}

I have presented a new method for computing time-dependent radiation
transport for an arbitrary distribution of sources embedded in an
arbitrary distribution of absorbing, emitting and scattering
material. The new algorithm, based on the MC radiative-equilibrium
method of \cite{lucy_1999} can be used in 3-dimensions and scales
almost perfectly under parallelization.  It has advantages over the
FLD approximation in that it is polychromatic and correctly treats the
directionality and flux of the radiation field in the optically thin
limit.  Note that although the algorithm as presented here is only applicable under conditions of LTE, the method may be straightforwardly extended to the  non-LTE regime e.g. \cite{carciofi_2007}.

I have applied the new method to the problem of a circumstellar disc
illuminated by a protostar with a periodic time-variable accretion
rate. I have shown that the lag between the blue continuum resulting
from the accretion hot spots and the reprocessed IR radiation from the
disc is a strong function of wavelength. It appears that photometric
time-series data in the blue part of the optical spectrum, combined
with equally intensive near-IR time-series could be used to probe the
geometrical and thermal structure of the disc, although it is likely
that complications arising from stochatic variations in the disc
structure close to the inner edge could mask the correlation
calculated here.

In future the time-dependent method will be incorporated in
radiation-hydrodynamics calculations of protostellar disc
fragmentation, extending the time-independent transfer simulations of
\cite{acreman_2010}.  It is well established that the thermal
properties of the disc strongly affect the fragmentation
\citep{boss_2008, stamatellos_2008}, and it is also clear that the FLD
approximation does not capture the full physics of the transport in
the disc itself.

\section*{Acknowledgements}

The calculations presented here were performed using the University of
Exeter Supercomputer. This research was in part funded by STFC grant
number ST/F003277/1. I thank the University of Exeter for the award of
study leave for the academic year 2009/2010.

\bibliographystyle{mn2e}
\bibliography{harries}
\end{document}